\documentclass[12pt,a4paper,preprintnumbers,superscriptaddress,showkeys,]{revtex4-1} %%% Use this class to prepare your letter for review

\usepackage[fleqn]{amsmath}
\usepackage{amssymb}
\usepackage{bm}% bold math
\usepackage{graphicx} % \includegraphics
\usepackage{subfigure}
\usepackage[pdfstartview=FitH,
            CJKbookmarks=true,
            bookmarksnumbered=true,
            bookmarksopen=true,
            colorlinks=true,
            pdfborder=001,
            linkcolor=blue,
            citecolor=blue,
            urlcolor=blue
            ]{hyperref}
\RequirePackage{color}
\newcommand{\scite}[1]{\textsuperscript{\cite{#1}}}  %%% \scite (Superscript citation with Square brackets) for the TAML and authors using RevTex 4.
\begin{document}

\preprint{Manuscript submitted to the \textit{Theoretical \& Applied Mechanics Letters}}

\title{Transition and self-sustained turbulence\\ in dilute suspensions of finite-size particles}
%\thanks{Footnote to title of article.}

\author{I. Lashgari}
\thanks{First Corresponding Author. Email: imanl@mech.kth.se}
\affiliation{Linn{\'e} FLOW Centre and SeRC, KTH Mechanics, Stockholm, Sweden}%%%
%\thanks{First Corresponding Author. Email: imanl@mech.kth.se}
\author{F. Picano}%
%\thanks{Email: Second.Author@institution.edu.cn}
\affiliation{Linn{\'e} FLOW Centre and SeRC, KTH Mechanics, Stockholm, Sweden}%%%
\affiliation{Industrial Engineering Department, University of Padova, Padova, Italy}%

%\author{Jiachun Li}
%\thanks{First Corresponding Author. Email: Third.Author.@institution.edu.}
%\affiliation{Editorial Office of Theoretical \& Applied Mechanics Letters, Beijing 100190, China}%
%\affiliation{\mbox{Editorial Board of Theoretical \& Applied Mechanics Letters, Beijing 100190, China}}%%%

\author{L. Brandt}
%\thanks{Second Corresponding Author. Email: luca@mech.kth.se}
\affiliation{Linn{\'e} FLOW Centre and SeRC, KTH Mechanics, Stockholm, Sweden}%%%

\date{\today}

\begin{abstract}
\noindent \textbf{Abstract} 

We study the transition to turbulence of channel flow of finite-size particle suspensions at low volume fraction, i.e. $\Phi \approx 0.001$. 
The critical Reynolds number above which turbulence is sustained reduces to $Re \approx 1675$,  in the presence of few particles, independently of the initial condition, a value lower than that of the corresponding single-phase flow, i.e.\ $Re\approx1775$.

In the dilute suspension, the initial arrangement of the particles is important to trigger the transition at a fixed Reynolds number and particle volume fraction. As in single phase flows, streamwise elongated disturbances are initially induced in the flow.   
If particles can induce oblique disturbances with high enough energy within a certain time, the streaks breakdown, flow experiences the transition to turbulence and the particle trajectories become chaotic. Otherwise, the streaks decay in time and the particles immigrate towards the channel core in a laminar flow. 
\\
\end{abstract}

%Use showkeys class option if keyword display desired

\maketitle

\section{Introduction}

Understanding the characteristics of suspension flows is of fundamental and practical importance in natural phenomena, e.g.\ particles in the atmosphere and water, and industry, e.g.\ transportation and mixing. The focus of this paper is therefore on the transition from the laminar to the turbulent flow of dilute suspensions of finite-size particles, particles larger than the smallest flow scales, a process associated to a significant (usually sudden) alteration of the nature of the flow. Although  the dynamics is governed by a single non-dimensional parameter -- the Reynolds number, the ratio of inertia to viscous forces, transition of single phase flows has challenged the scientists for a long time and it is not yet completely understood. The behavior of suspensions is more complicated because of the various particle properties such as size, number, shape, deformability, density. 
  
To the best of our knowledge, there exist only few studies of the transition to turbulence of suspensions of finite-size particles (for the case of point particles the reader is referred to \cite{joy} and references therein). The experiments by Matas \emph{et. al.}\scite{Matas03} examine the effects of finite-size neutrally buoyant particles on the transition in pipe flow. These authors report that suspensions of large particles exhibit a non-monotonic behavior of the critical Reynolds number 
when increasing the particle volume fraction. The different regimes are identified  by the pressure drop between the inlet and outlet of the pipe. A decade later and thanks to the improvement of computational algorithms and resources, numerical simulations of finite-size particle suspensions start to emerge. Yu et.\ al.\scite{Yu13} partially simulate the experiments  in \cite{Matas03}. Since the flow is always perturbed by the presence of the particles, the level of streamwise velocity fluctuations is used to define a threshold to distinguish between laminar and turbulent flow. The experimental behavior in \cite{Matas03} could be reproduced by tuning this threshold parameter, showing the difficulties to define the transition threshold in suspensions. A more detailed analysis of the flow in the transitional regime is reported by Loisel et.\ al.\scite{Loisel13} where a fixed particle volume fraction is examined, $\Phi \approx 5\%$. These authors show that the coherent structures of the flow are broken by the presence of finite size particles and smaller eddies (more energetic) prevents the flow from relaminarization when decreasing the Reynolds number;  this effect promotes therefore turbulence. 

Summarizing, transition delay is attributed to the enhancement of the effective viscosity of the suspensions for smaller particles \scite{Matas03}, 
whereas promotion of transition is, instead, qualitatively attributed to large disturbances 
induced by particles of large enough size. 

Recently Lashgari et.\ al.\scite{Lashgari14} 
studied suspensions of spherical neutrally buoyant particles for a wide range of Reynolds numbers, $Re$, and box-averaged volume fractions, $\Phi$. These authors examine  the global momentum balance \scite{Zhang10} and report the existence of three different regimes when varying $\Phi$ and $Re$.  For low $\Phi$ and $Re$, the flow is laminar and the viscous stress dominates. For high Re and sufficiently low $\Phi$, the flow is turbulent and the Reynolds stress contributes the most to the momentum transport as in classic single-phase turbulence. The flow is dominated by the particle stress at moderate $\Phi$.

For the cases at low $\Phi$, transition is sharp when increasing the Reynolds number and can be easily identified, e.g.\ by the level of fluctuations and wall shear stress; at high $\Phi$, however, all the observables vary smoothly with $Re$. The latter case is denoted as inertial shear-thickening since it is characterized by a significant increase of the wall friction that is not attributed to an increase of the Reynolds stress but to the enhancement of the particle-induced stress. 

The aim of this letter is to examine 
the reduction of the critical Reynolds number above which turbulence exists in very dilute suspensions of finite size particles.
We study in details the mechanism behind the transition promotion and relate this to self-sustained turbulence by analysing the kinetic energy induced by the particles and transferred from small scales to large scales. 
This paper is organized as follows. In $\S II$, we introduce the numerical method and the simulation setup. The results are presented and discussed in  $\S III$; a summary of the main conclusion is presented in  $\S IV$.

\section{numerical method and problem setup}
We perform direct numerical simulation of suspensions laden with rigid spherical neutrally buoyant particles. We employ an immersed boundary solver based on the original formulation by Uhlmann\scite{Uhlmann05} and developed by Breugem\scite{Breugem12}. The code couples a fixed uniform Eulerian mesh for the fluid phase with a quasi-uniform Lagrangian mesh representing the surface of the particles. The fluid velocity is interpolated on the Lagrangian grids, the immersed boundary forcing is computed based on the difference between the particle velocity and the interpolated fluid velocity at each Lagrangian grid point and finally the forcing spread out from the Lagrangian to the Eulerian mesh. The near field interactions are treated by means of lubrication forces and soft-sphere collision models. The code has been validated against several test cases in \cite{Breugem12,Lashgari14}. 

We simulate the flow in a pressure-driven channel flow with streamwise and spanwise periodic boundary condition and no slip condition at the walls. The box size is $2h\times3h\times6h$ in the wall-normal, spanwise and streamwise directions where $h$ is the half channel height. The domain is larger than the minimal unit channels used for transition in Newtonian fluids\scite{Hamilton95} and polymer suspensions\scite{ Graham10}. The number of Eulerian grid points is $160\times240\times480$ with 746 Lagrangian points used to resolve the surface of each particle. The ratio between the channel height and particle dimeters is fixed to 10 (with 16 grid point per particle diameter). The particle diameter is that pertaining the case in the experiment\scite{Matas03} where the strongest non-monotonic behavior of the critical transition threshold is observed. We denote streamwise coordinate and velocity by $v$ and $y$, wall normal by $w$ and $z$ and spanwise by $x$ and $u$. The simulations are performed imposing a constant mass flux, with the bulk velocity denoted by $U_b$.  The Reynolds number is defined as $Re=2U_bh/\nu$ where $\nu$ is the fluid kinematic viscosity. 
In order to calculate the characteristics of the two-phase flow, a phase field indicator (mask), $\xi$, is created for the total field such that  $\xi= 1$ indicates the solid phase and $\xi = 0$ the fluid phase.  The parameters of the fluid phase, e.g. rms velocities,  are then obtained by taking average over all the Eulerian points with $\xi = 0$ and similar for the particle phase when $\xi = 1$.

\section{results}
In this work, we study the transition and self-sustained turbulence in a channel flow laden with few finite-size neutrally-buoyant particles (dilute regime) and compare the results with the one of the single-phase (Newtonian) flow. We use only 10 particles corresponding to a particle volume fraction $\Phi\approx0.001$.

\subsection{Threshold of sustained turbulence}

\begin{figure}[t]
\includegraphics[width=0.45\linewidth]{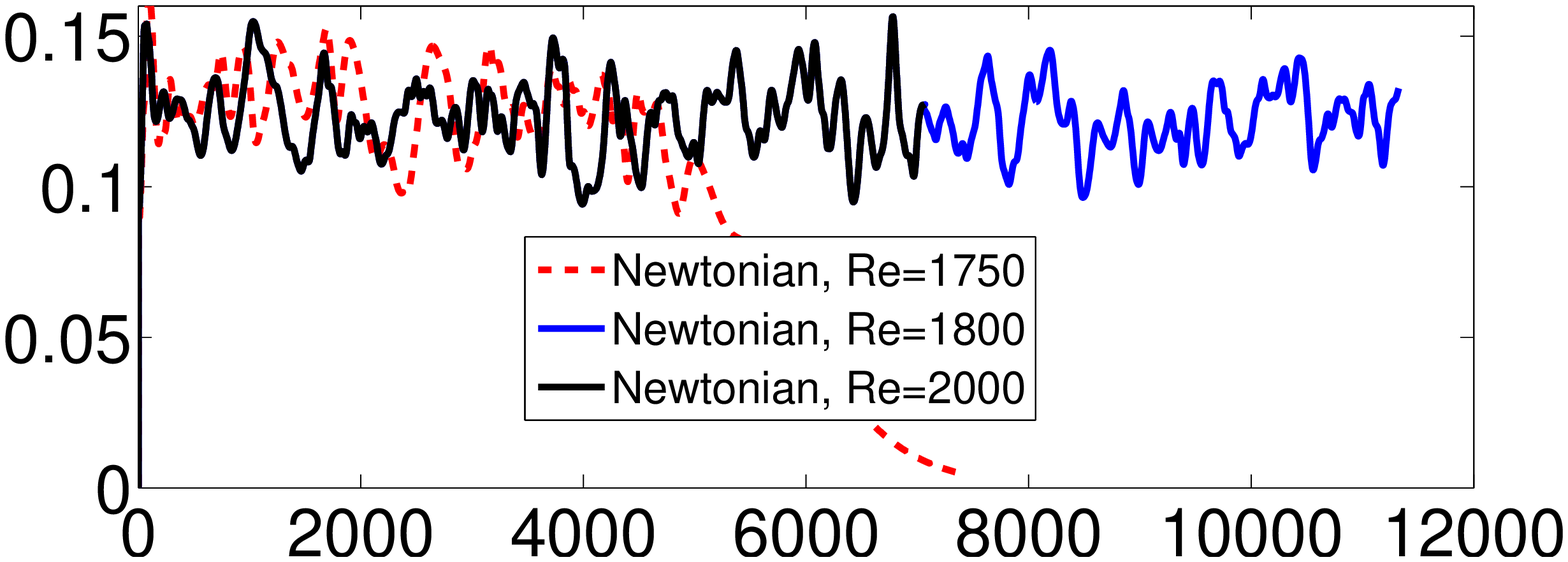}
\put(-210,75){{\large $(a)$}}
\put(-220,40){${v_{rms}}$}
\put(-110,-10){{$t$}}
\includegraphics[width=0.45\linewidth]{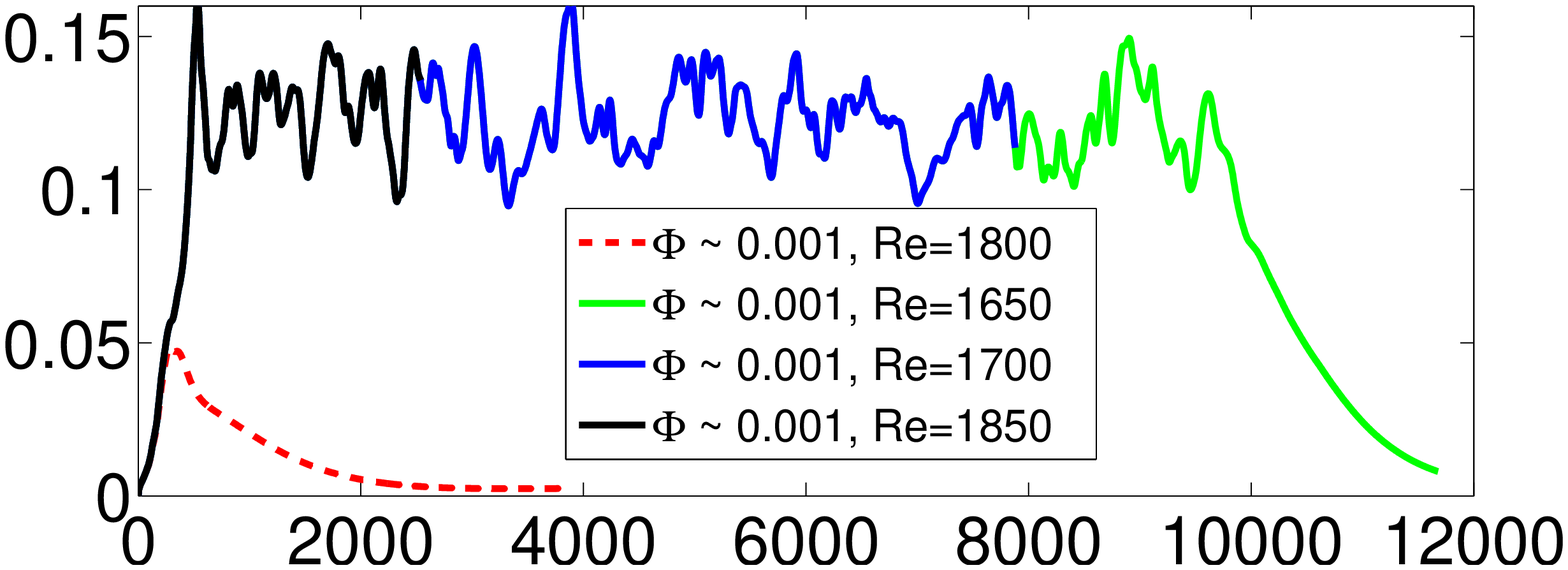}
\put(-210,75){{\large $(b)$}}
\put(-220,40){${v_{rms}}$}
\put(-110,-10){{$t$}}
\\
\includegraphics[width=0.45\linewidth]{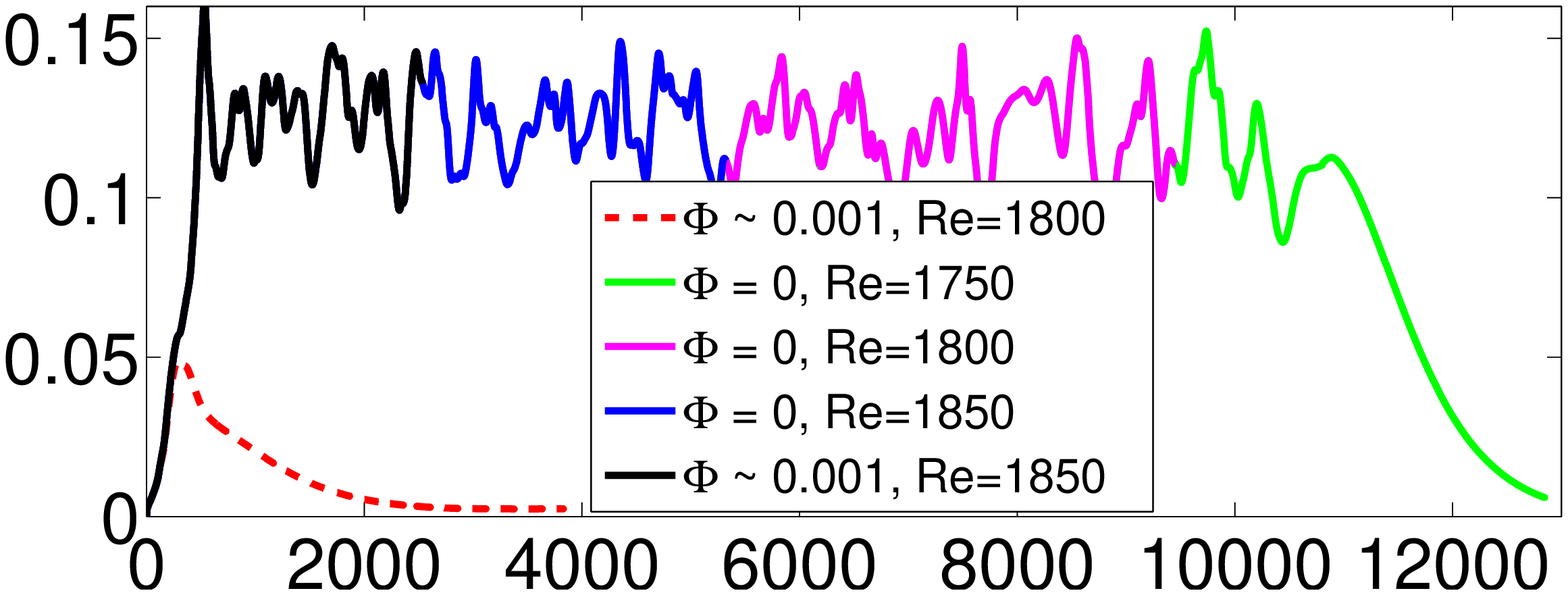}
\put(-210,75){{\large $(c)$}}
\put(-220,40){${v_{rms}}$}
\put(-110,-10){{$t$}}
\caption{\label{fig:vrms} Time history of the streamwise velocity fluctuations for a) Newtonian flow, b) and c) two different paths in particle laden flow (See text).}
\end{figure}

It is known that the transition threshold in Poiseuille flow depends on the initial disturbance; one strategy to obtain the threshold for sustained turbulence is to decrease the Reynolds number of the turbulent flow until the flow re-laminarizes  (see among others \cite{Eckhardt00,Philip11}). 
For the single phase flow, we use as initial disturbance high amplitude localized stream-wise vortices; see \cite{Henningson91} for the analytical expression of the disturbance velocity field.  
The time histories of the streamwise rms (root mean squared) perturbation velocity of the unladen flow are depicted in figure \ref{fig:vrms}(a). Note that the rms velocities are normalized by $U_b$ and time in units of $h/U_b$. The critical Reynolds number 
is found to be $1750<Re_c<2000$ for this particular flow domain and a maximum wall normal velocity of the initial disturbance equal to 10 times the bulk velocity;  fluctuations are sustained at $Re=2000$ but they decay and eventually vanish at $Re=1750$. 
In the next step, we initiate the simulations with a turbulent velocity field at $Re_b=2000$, decrease the Reynolds number to $1800$ and run the simulation for a long time: the fluctuations remain and therefore the threshold value of the sustained turbulence of the unladen flow is identified approximately, $Re_c\approx 1775$.

The initial condition for the particle-laden flow is given by a random arrangement of the particles, all moving with the local fluid velocity and  initial angular velocity equal to half the value of the local vorticity. 
The initial disturbance source is due to the flow adjustments to the particle presence
and therefore depends on the particle position. 
We initially keep the same initial condition and run simulations at different Reynolds numbers. We observe that the fluctuations, induced by the particles, eventually decay at $Re<1800$ while they grow to the turbulent regime at $Re>1850$ (see the fluid streamwise rms velocity in figure \ref{fig:vrms}b).  Note that the initial particle arrangement strongly affects the disturbance growth and transition. Therefore, the threshold value of $1800<Re_c<1850$ is only valid for this particular initial configuration; a different behavior is most likely to be observed with another random initial distribution. We will come back to this point when we analyze the trajectory of the particles. As for the unladen flow, we therefore reduce the Reynolds number of the particulate turbulent flow and monitor the Reynolds number at which flow re-laminarizes. The results in the figure reveal that turbulence is sustained at $Re=1700$ while it decays at $Re=1650$. Based on these set of simulations, the critical value for sustained turbulence in the particle laden flow is $Re \approx 1675$, a value lower than that of the unladen flow in agreement with the experimental data by Matas \emph{et. al.} \scite{Matas03}. 
The finite-size particles do not only trigger the transition to turbulence but also keep the turbulence at Reynolds number lower than that of the unladen flow. Note that the simulations at which turbulence is sustained are integrated for a time longer than that shown in figure  \ref{fig:vrms}.

To further highlight the difference between laden and unladen flow, we perform a third set of simulations, see figure \ref{fig:vrms}c). In this case 
starting with a turbulent flow  at $Re=1850$, we first remove the particles and then reduce the Reynolds number until the flow re-laminarizes. The flow sustains turbulence at $Re=1800$ and it re-laminarizes at $Re=1750$ providing additional evidence that the threshold for unladen flow is about $Re\approx1775$ and that the presence of particles is important to sustain turbulence. For all the cases studied, the fluctuations decay at similar rate once the flow re-laminarizes.   We have compared the statistics of the turbulent laden and unladen flows at the same Reynolds number and observe small differences (not shown here).

\subsection{Particle trajectories}

The initial particle configuration  
plays a vital role to determine the final state of the flow.  
The flow can become either 
laminar or turbulent at fixed Reynolds number and number of particles for different initial arrangement of the particles. 

To gain physical understanding on the particle influence, we compare the flows at $Re=1850$ $\&$ $\Phi \approx 0.001$ and $Re=1800$ $\&$ $\Phi \approx 0.001$ and the same initial random arrangement of the particles (denoted as case-I and case-II), with a simulation at $Re=1850$ $\&$ $\Phi \approx 0.001$ and a different initial random arrangement (case-III). 
Out of these three flows, only case-I is turbulent whereas the other  two are laminar. Although the threshold for sustained turbulence for the particle-laden flow is $Re \approx 1675$, the disturbances produced by the particles may not be strong enough to bring the system to turbulence. 

\begin{figure}[t]
\includegraphics[width=0.45\linewidth]{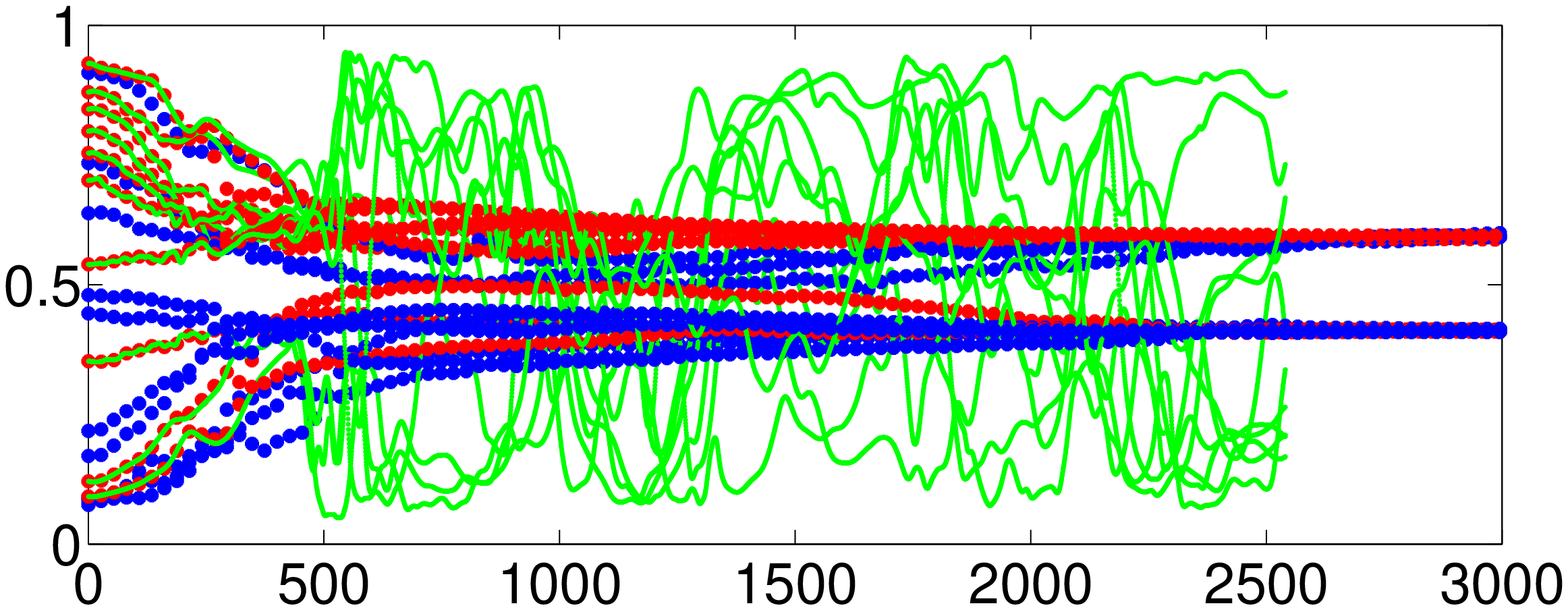}
\put(-215,70){{\large $(a)$}}
\put(-205,40){{${z}$}}
\put(-120,-10){{$t$}}
\includegraphics[width=0.45\linewidth]{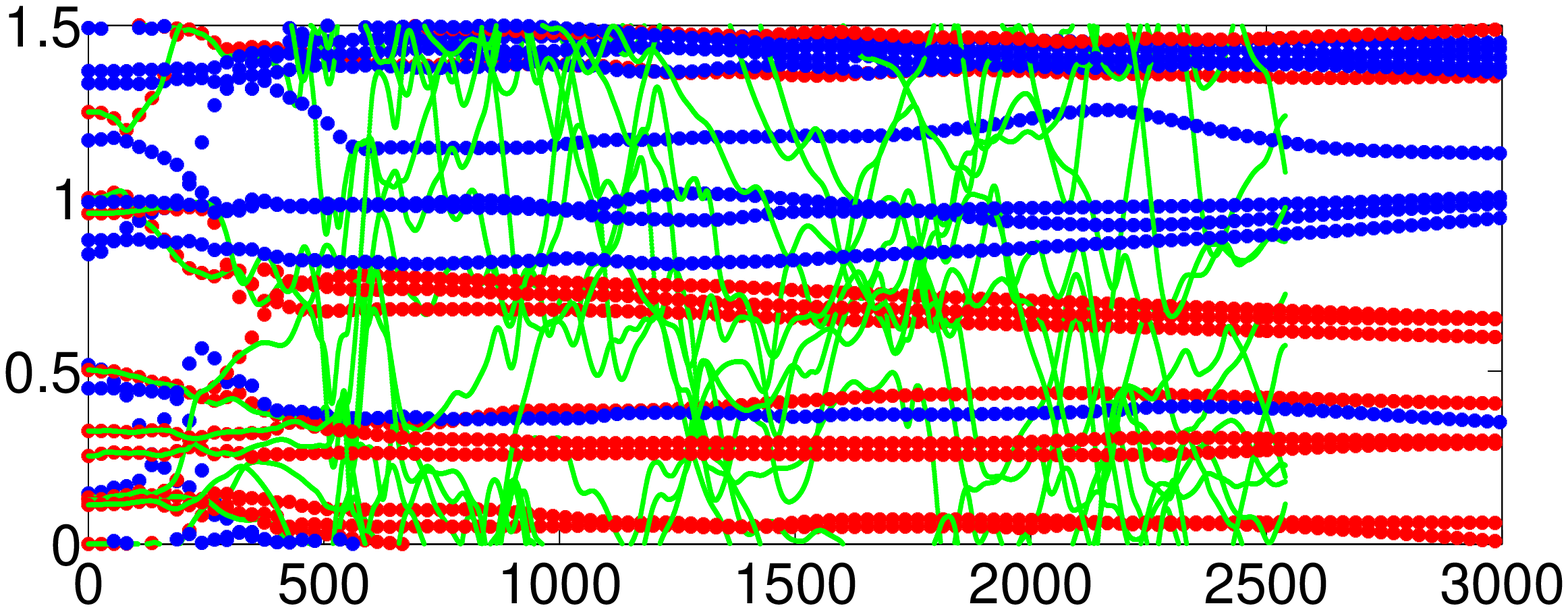}
\put(-215,70){{\large $(b)$}}
\put(-205,40){{${x}$}}
\put(-120,-10){{$t$}}
\\
\includegraphics[width=0.45\linewidth]{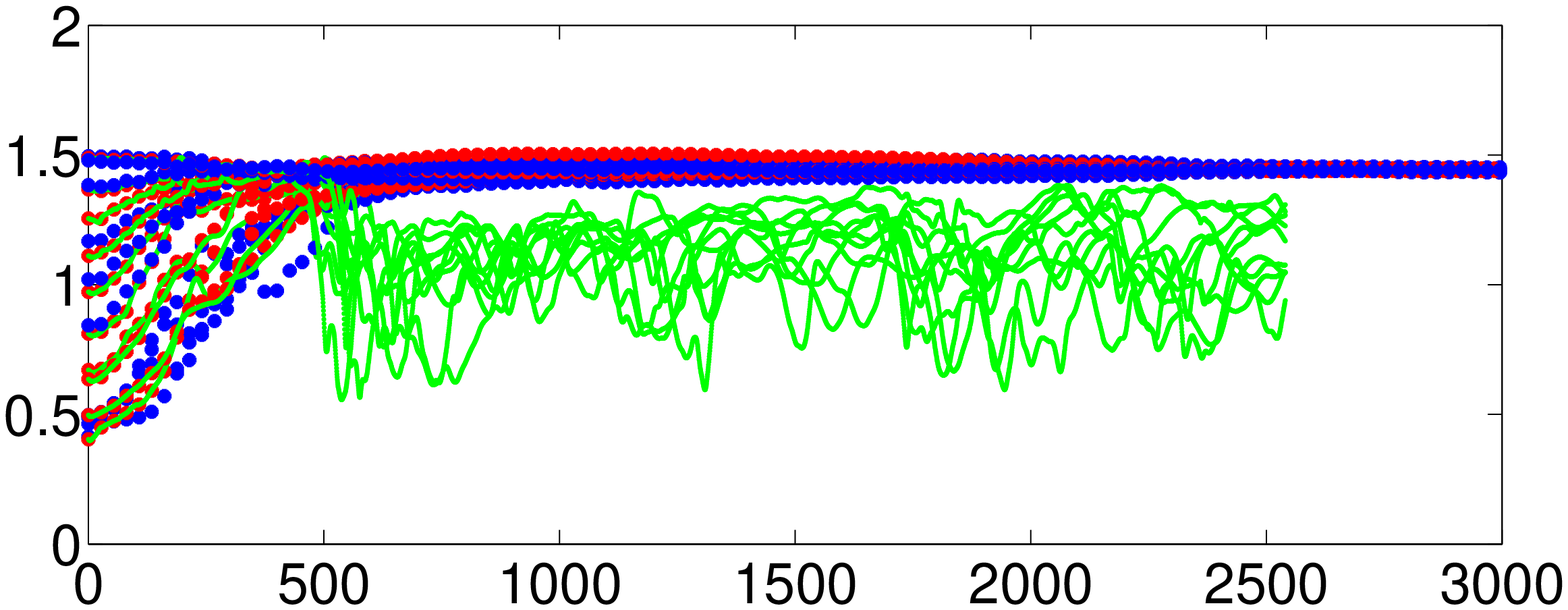}
\put(-215,70){{\large $(c)$}}
\put(-205,40){{${v}$}}
\put(-120,-10){{$t$}}
\caption{\label{fig:trajectory} Particle trajectories in: (a) wall normal and (b) spanwise  directions, and c) particle streamwise velocity for case-I (green symbols), case-II (red symbols) and case-III (blue symbols). See text for the definition.}
\end{figure}

The time histories of the wall-normal particle position are reported in  figure \ref{fig:trajectory}(a) for the three cases introduced above.
The particles of case-I (green symbols) and case-II (red symbols), with same initial distribution, initially follow the same path. 
At $t \approx 300$, the trajectories start to deviate and exhibit a completely different behavior at $t > 500$. Case-I is characterized by a  chaotic particle motion, whereas particles migrate toward the center of the channel in case-II; they are eventually found at about $10\%$ of the channel height, from either side of the channel centerline, $z\approx 0.45\,- 0.55$.
This migration is explained by the action of the Saffman lift force (see Saffman \scite{Saffman65}  and Segre and Silberberg \scite{Segre61}). 
A similar behavior is observed for case-III (blue symbols) when the flow returns to the laminar  state although $Re=1850$. 
A detailed analysis of the particle final equilibrium position in the presence of inertia and as a function of particle size can be found in the work by Matas \emph{et. al.} \scite{Matas04} for laminar flows. 

We display the particle trajectories in the spanwise directions in figure \ref{fig:trajectory}(b). For the laminar cases (case-II and case-III), the trajectories are almost spanwise independent: particle are transported by the laminar fluid and do not experience strong lateral motions. Significant lateral motions occur, on the contrary, in a turbulent flow. The streamwise particle velocity is shown in figure \ref{fig:trajectory}(c). Here we observe a final streamwise velocities of about $1.44$ at the equilibrium position in the laminar cases, a value larger than the mean velocity (about 1.12) of the turbulent case.

We also examine 
the particle trajectories in suspensions of higher volume fractions and observe a similar final arrangement of the particles when the flow remains laminar. In other words, particles perturb the flow while migrating towards their equilibrium position.
If the flow becomes turbulent during this transient phase, particles will be subject to strong hydrodynamics forces and start to move chaotically; thus, in turn, they contribute to maintain the turbulence.

\subsection{Energy spectra}

\begin{figure}[t]
\includegraphics[width=0.33\linewidth]{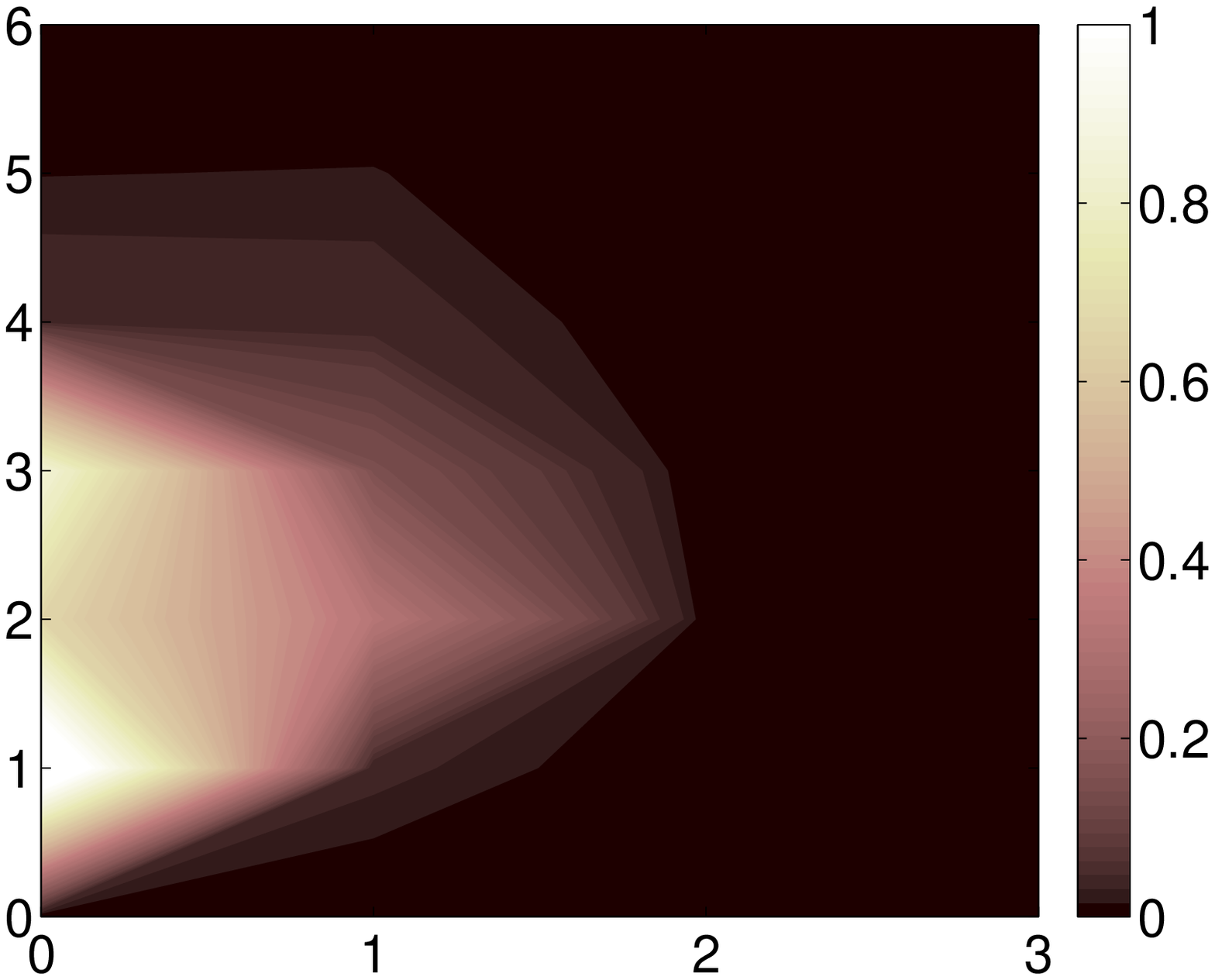}
\put(-140,120){{\large $(a)$}}
\put(-150,60){{${\beta}$}}
\put(-80,-0){{$\alpha$}}
\put(-100,115){{$t=300$}}
\includegraphics[width=0.33\linewidth]{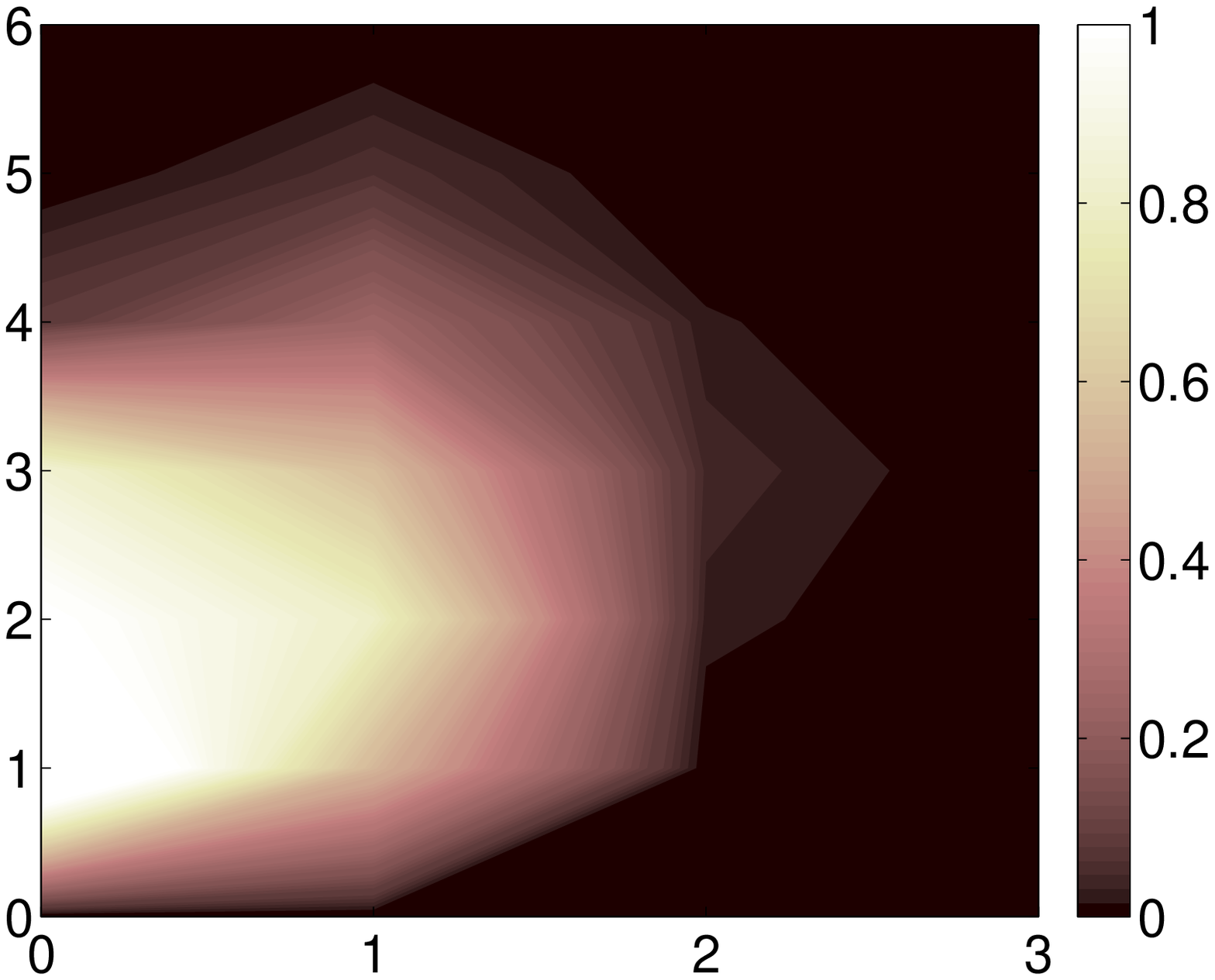}
\put(-140,120){{\large $(b)$}}
\put(-150,60){{${\beta}$}}
\put(-80,-0){{$\alpha$}}
\put(-100,115){{$t=400$}}
\includegraphics[width=0.33\linewidth]{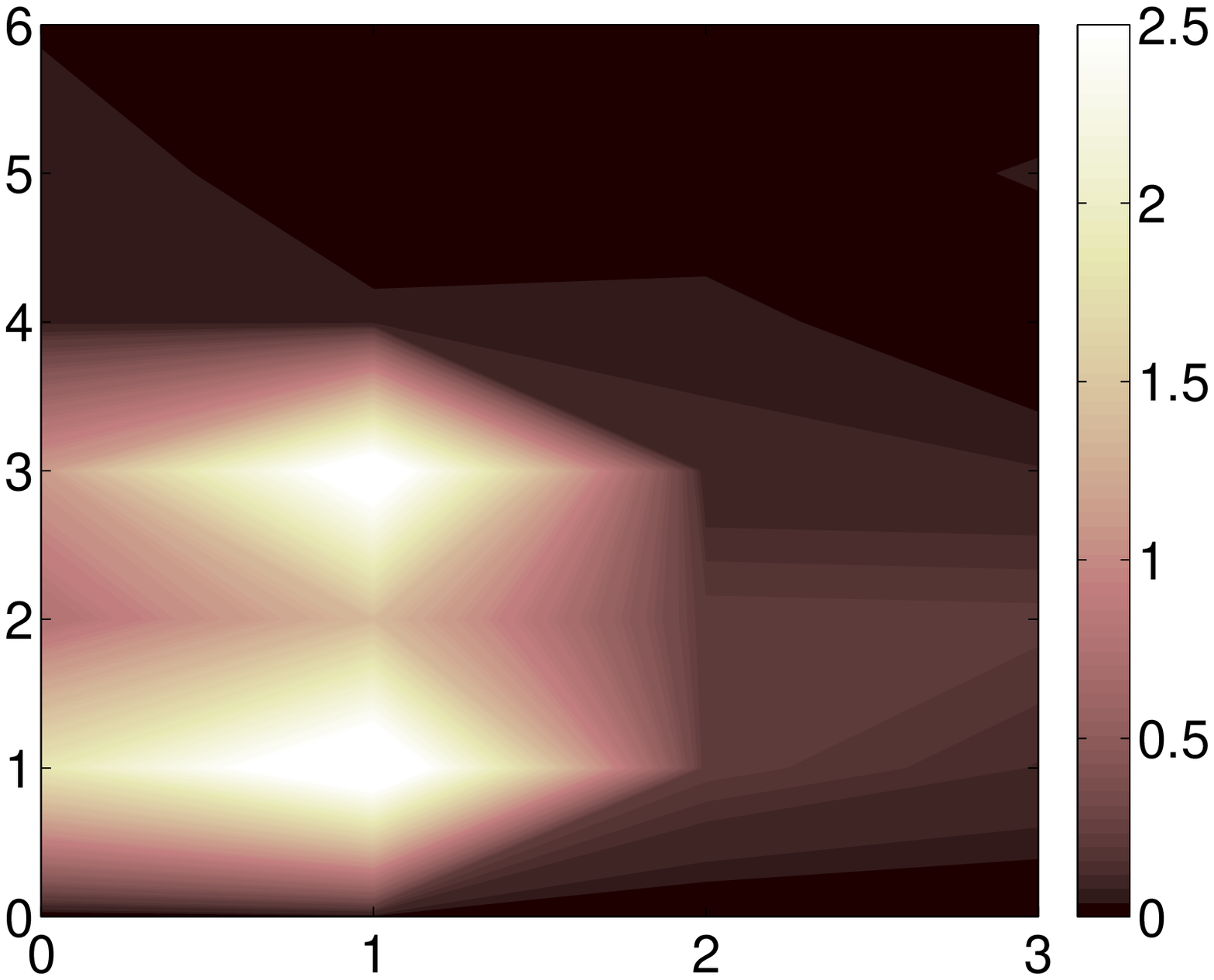}
\put(-140,120){{\large $(c)$}}
\put(-150,60){{${\beta}$}}
\put(-80,-0){{$\alpha$}}
\put(-100,115){{$t=500$}}
\\
\includegraphics[width=0.33\linewidth]{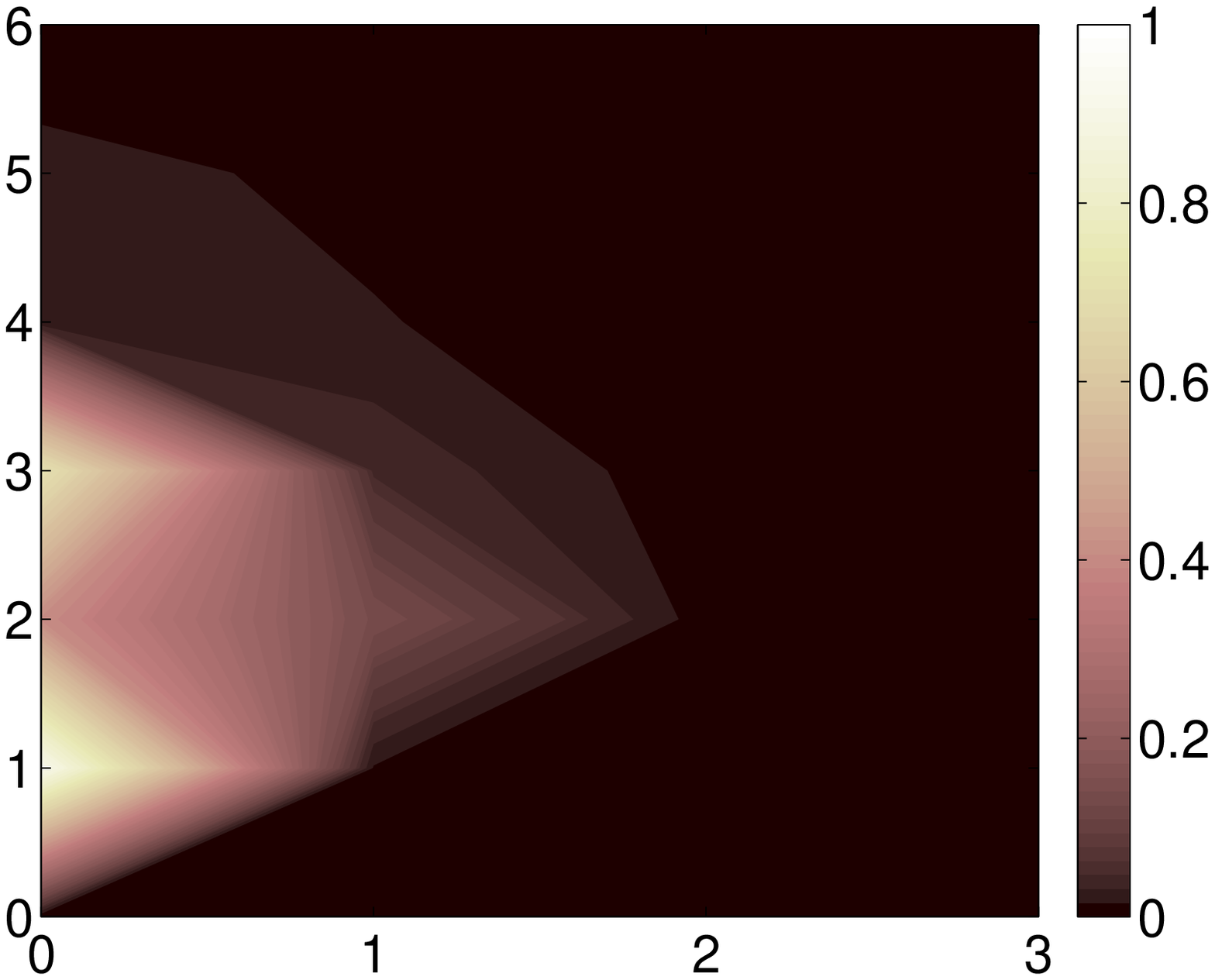}
\put(-140,120){{\large $(d)$}}
\put(-150,60){{${\beta}$}}
\put(-80,-0){{$\alpha$}}
\put(-100,115){{$t=300$}}
\includegraphics[width=0.33\linewidth]{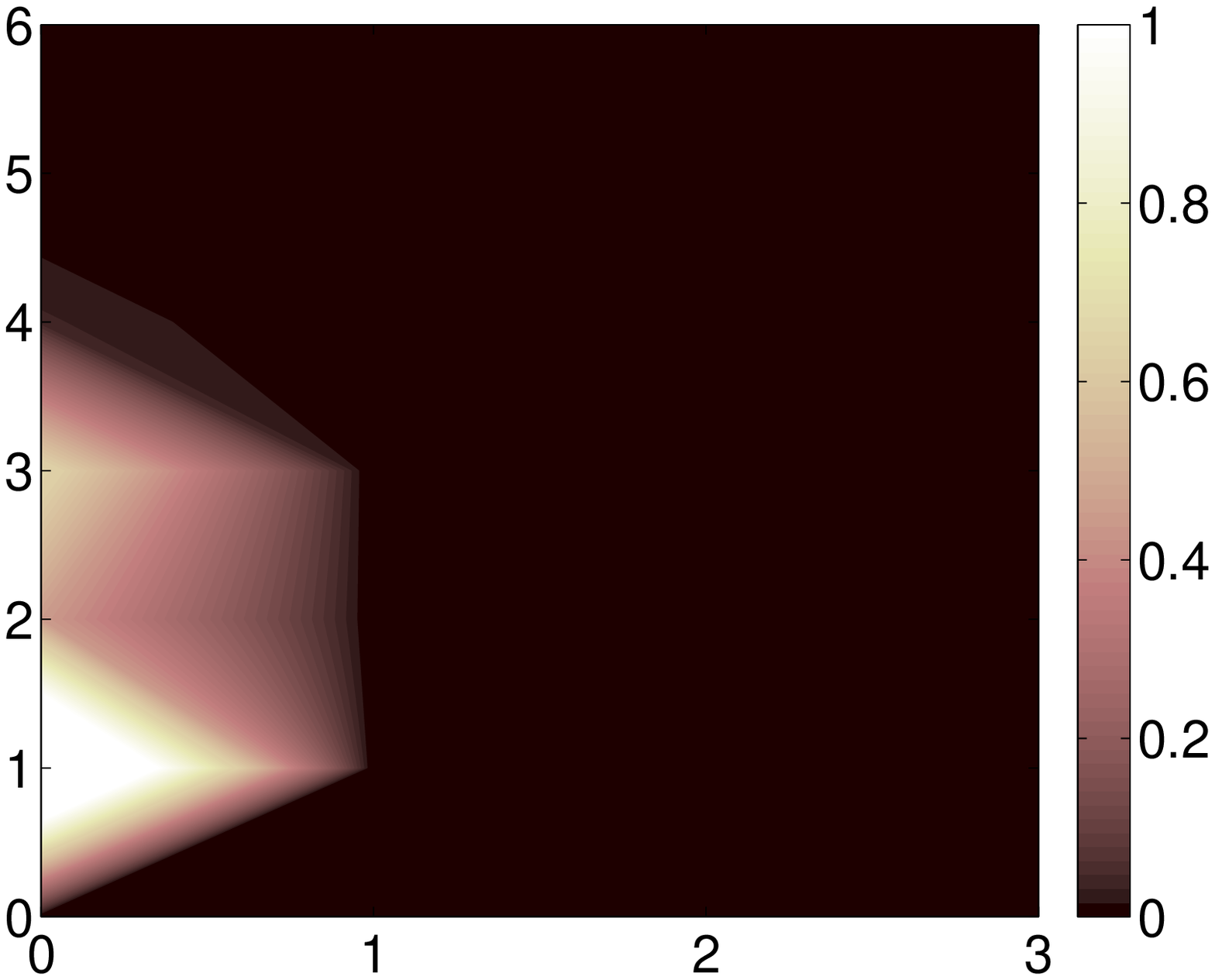}
\put(-140,120){{\large $(e)$}}
\put(-150,60){{${\beta}$}}
\put(-80,-0){{$\alpha$}}
\put(-100,115){{$t=400$}}
\includegraphics[width=0.33\linewidth]{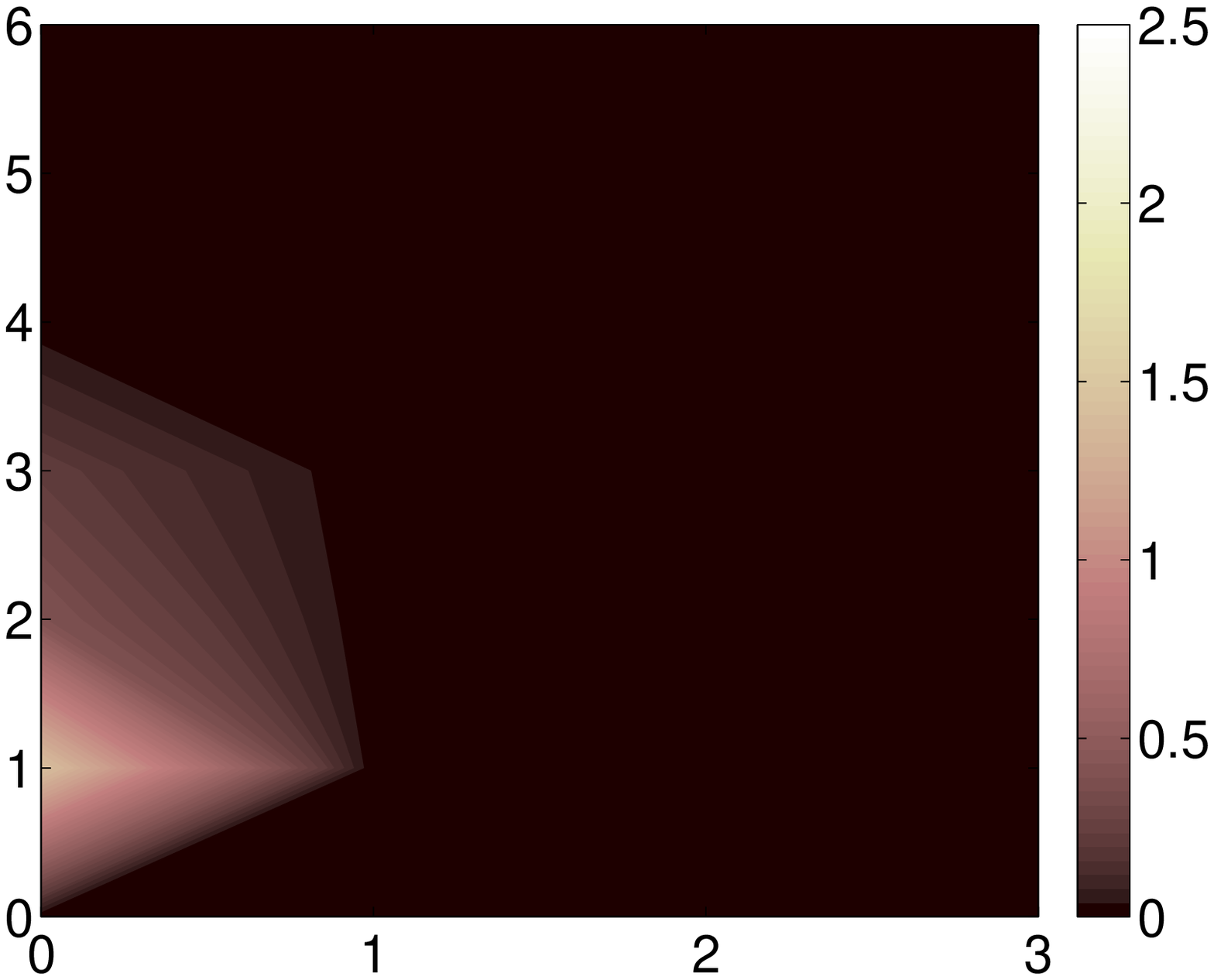}
\put(-140,120){{\large $(f)$}}
\put(-150,60){{${\beta}$}}
\put(-80,-0){{$\alpha$}}
\put(-100,115){{$t=500$}}
\\
\includegraphics[width=0.33\linewidth]{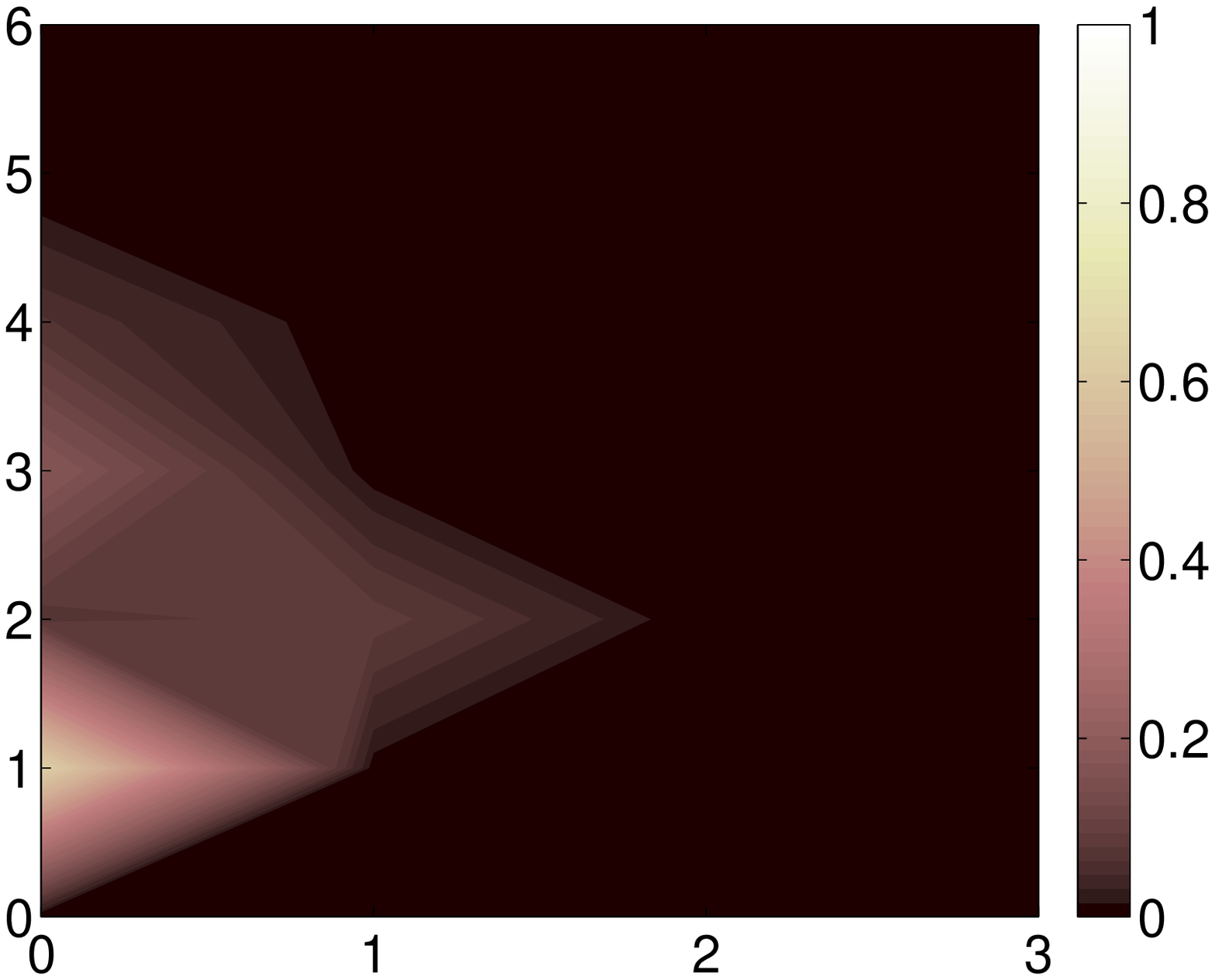}
\put(-140,120){{\large $(g)$}}
\put(-150,60){{${\beta}$}}
\put(-80,-0){{$\alpha$}}
\put(-100,115){{$t=300$}}
\includegraphics[width=0.33\linewidth]{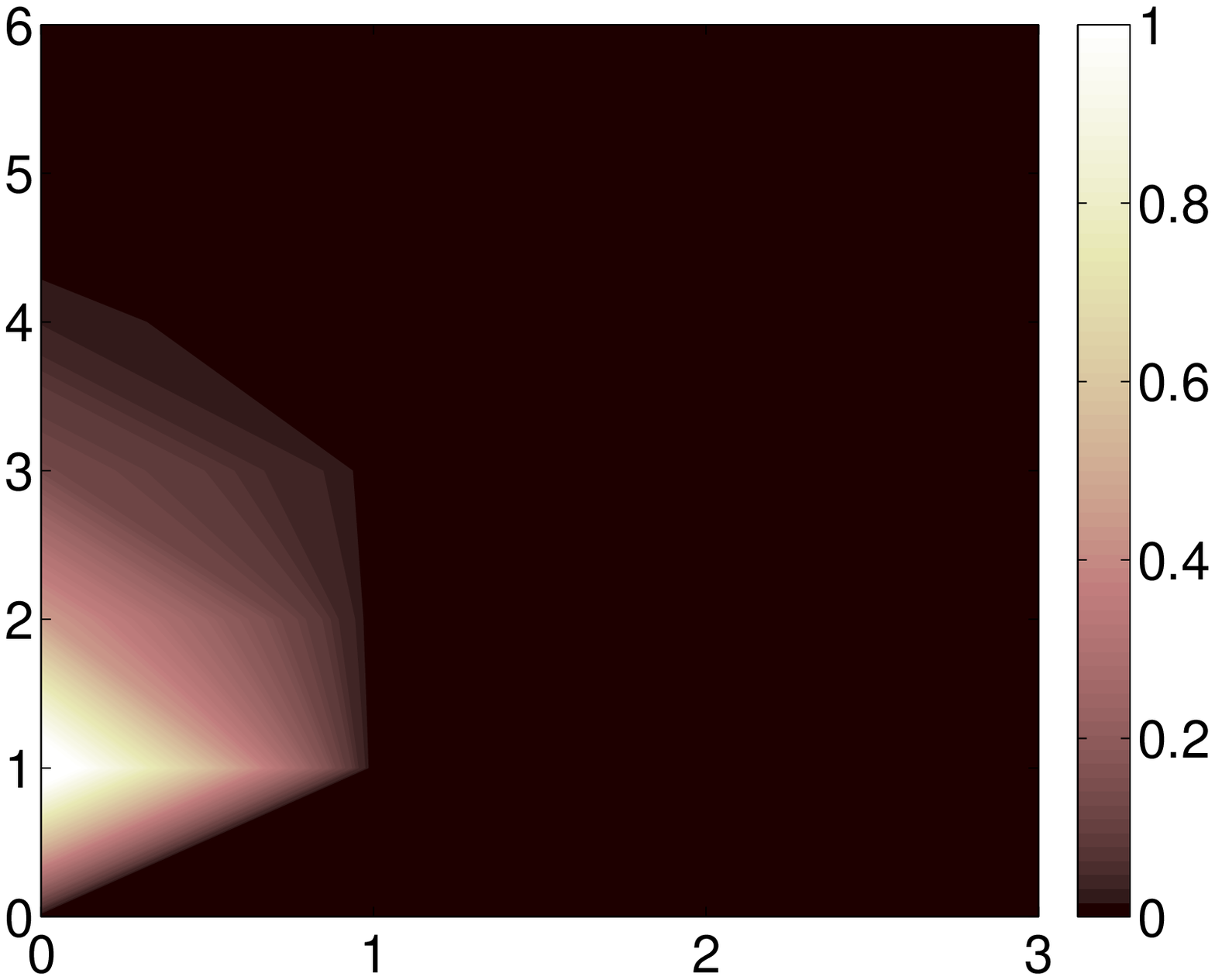}
\put(-140,120){{\large $(h)$}}
\put(-150,60){{${\beta}$}}
\put(-80,-0){{$\alpha$}}
\put(-100,115){{$t=400$}}
\includegraphics[width=0.33\linewidth]{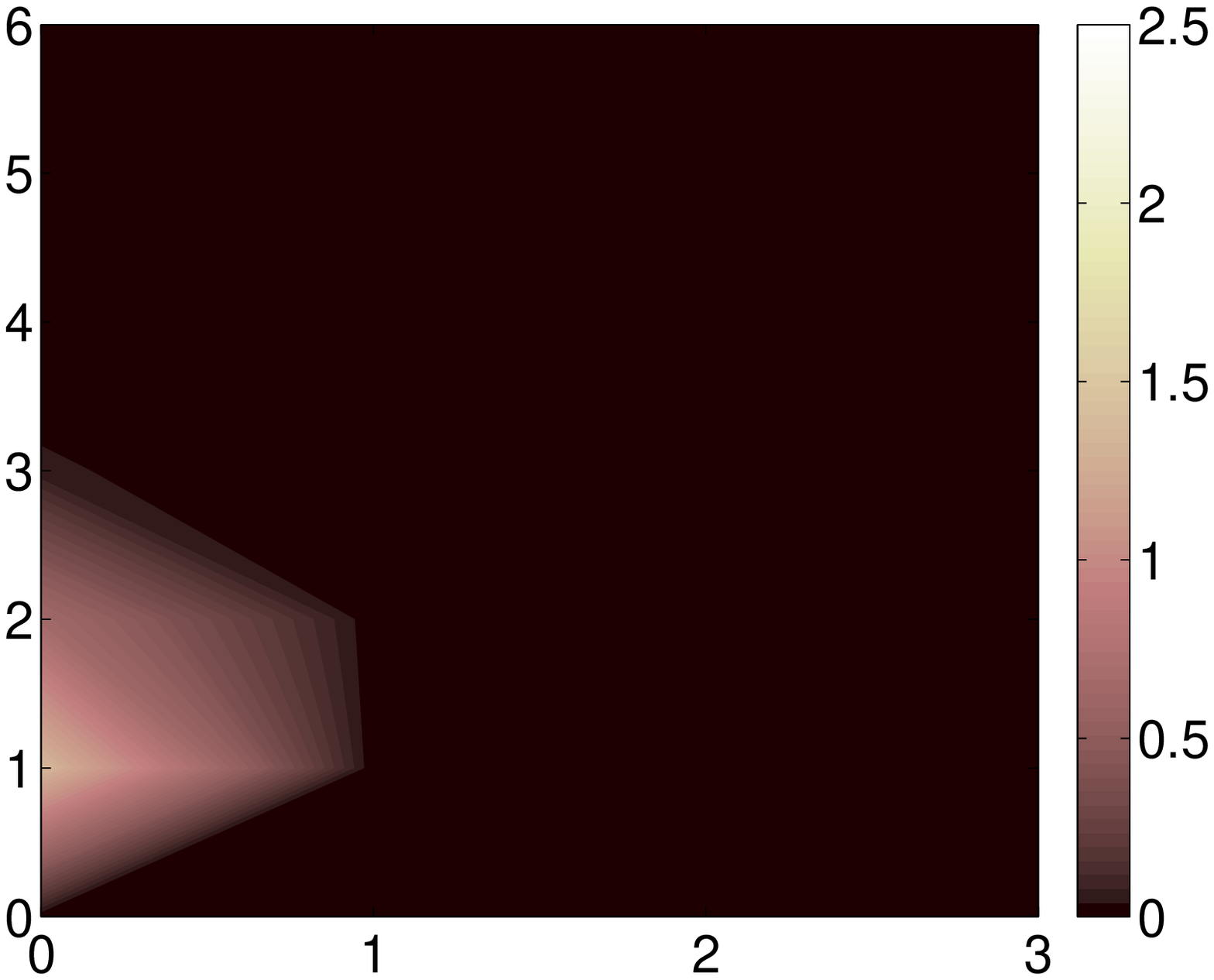}
\put(-140,120){{\large $(i)$}}
\put(-150,60){{${\beta}$}}
\put(-80,-0){{$\alpha$}}
\put(-100,115){{$t=500$}}
\caption{\label{fig:spectra} Energy spectra for the flow case-I (a,b,c) , case-II(d,e,f) and case3 (g,h,i) and (a,d,g): $t= 25$, (b,e,h): $t = 300$, (c,f,i): $t = 500$}
\end{figure}

In order to better understand the mechanism behind the transition in the presence of few particles, we examine the two-dimensional spectra of the perturbation kinetic energy, integrated in the wall-normal direction. We shall denote by $(\alpha,\beta)$ the streamwise and spanwise wave numbers. The number of Fourier modes employed for each case is enough to reproduce the original field with less than $1\%$ error. 

We report the energy spectra pertaining case-I, case-II and case-III in figure \ref{fig:spectra} at three different times in the interval when the behavior of the flow changes considerably,
$ t\in[300, 500]$,  (see figure \ref{fig:trajectory}). 
The contours on the top/middle/low panel correspond to case-I/case-II/case-III. The energy spectra at early times clearly reveal that  particles contribute to introduce energy to small scales (not shown here); the modes with high streamwise and spanwise wavenumbers contain non-negligible energy when particles are introduced in the flow. 
By the time $t=300$, the energy is transferred back to the larger scales for all cases (see figure \ref{fig:spectra}(a,d,g)). The peak of the spectra are located at $\alpha=0$, corresponding to the formation of elongated streaks by the lift-up effect. 
As discussed in Brandt \scite{Brandt14}, the linear lift-up effect, responsible for the streak formation, is hardly affected by the particle presence. Once the streaks have formed, the difference between the three cases is in the energy content of  modes with $\alpha \ge 1$: this is found to be larger for case-I, the only turbulent flow. This is likely due to the particular initial condition (case-I vs.\ case-III) and to the slightly larger Reynolds numbers (case-I vs.\ case-II).
This is more evident at $t=400$, figure \ref{fig:spectra}(b),  where oblique modes are strong enough to destroy the streaks and promote transition. The flow becomes streamwise dependent and the disturbance energy cascades to modes of higher and higher wavenumbers, see figure  \ref{fig:spectra}(c). The opposite is true for case-II and case-III, figure  \ref{fig:spectra}(e,f,h,i): here a single streak resists against the perturbations induced by the particles until it eventually decays. 
Examining the evolution of the perturbation kinetic energy,  we thus infer that the particles induce streamwise dependent modes (oblique waves). If these oblique modes have high enough energy, they promote the streak breakdown and the flow undergoes the transition to turbulence. Conversely, the streaks decay and
particles migrate towards their equilibrium position, close to the channel centerline. The initial arrangement of the particles is thus directly connected to the streak breakdown and transition. In the future, it would be interesting to investigate the relative position between  vortices/streaks  and the particles to access a mechanistic model of their interactions.

\section{Conclusion and remarks}

We study the transition to turbulence in channel flow of dilute suspensions of finite-size neutrally buoyant particles and compare the results with those of the single-phase flow. The particle volume fraction studied is $\Phi \approx 0.001$.  An immersed boundary solver is used to simulate the particulate flow where lubrication forces and soft sphere collision models are implemented for the near field interactions. 

The critical Reynolds number above which turbulence is sustained is  $Re\approx1775$ for the single-phase flow. 
It decreases to $Re_c \approx 1675$ for the particulate flow with $\Phi \approx 0.001$. The disturbances induced by the particles are therefore enough to sustain turbulence at lower Reynolds numbers. The same threshold for the single-phase flow is also obtained, $Re\approx1775$, by removing the particles from a turbulent particle-laden flow and decreasing the Reynolds number. 

We show that the initial random arrangement of the particles is important to determine whether the flow becomes turbulent. If the flow undergoes transition to turbulence, the particles move chaotically whereas they migrate towards their equilibrium position, close to the centerline, in a laminar flow. This  is also observed for flows with a higher number of particles (e.g.\ 50 particles, $\Phi \approx 0.005$).   

We further examine the two-dimensional energy spectra
for
three cases; turbulent case-I ($Re=1850$)  and laminar case-II ($Re=1800$) with the same initial random arrangement of particles and laminar case-III ($Re=1850$) but with a different initial arrangement. The results indicate that the transition in case-I is due to higher energy content in  oblique disturbance modes. The particles can trigger strong enough oblique modes that promote the streak breakdown. The streaks undergo secondary instabilities and breakdown to turbulence, before being regenerated by non-linear interactions in a self-sustaining cycle\scite{Hamilton95}.  
In the other two cases, the streaks induced initially by the lift-up effect decay slowly in time and the particles migrate towards the centerline.

%%%%%%%%%%%%%%%%%%

%%%%%%%%%%%%%%%%%

%%%%%%%%%%%%%%%%
%\bibliographystyle{aipnum4-1}
%\bibliography{biblo}% Produces the bibliography via BibTeX.
%%%%%%%%%%%%%%%

\end{document}